\documentclass{ws-ijmpa}
\usepackage[super,compress]{cite}
\usepackage{graphicx}
\usepackage{color}

\newcommand{\be}{\begin{equation}}
\newcommand{\ee}{\end{equation}}
\newcommand{\bea}{\begin{eqnarray}}
\newcommand{\eea}{\end{eqnarray}}
\newcommand{\beq}{\begin{equation}}
\newcommand{\eeq}{\end{equation}}
\newcommand{\nn}{\nonumber}

\def\la{\mathrel{\mathpalette\fun <}}
\def\ga{\mathrel{\mathpalette\fun >}}
\def\fun#1#2{\lower3.6pt\vbox{\baselineskip0pt\lineskip.9pt
\ialign{$\mathsurround=0pt#1\hfil##\hfil$\crcr#2\crcr\sim\crcr}}}

\begin{document}

\title{PROTON-PROTON, PION-PROTON AND PION-PION DIFFRACTIVE COLLISIONS
 AT ULTRA-HIGH ENERGIES}

\author{V.V. ANISOVICH}
\address{National Research Centre "Kurchatov Institute",
Petersburg Nuclear Physics Institute, Gatchina 188300, Russia}

\author{K.V. NIKONOV}
\address{National Research Centre "Kurchatov Institute",
Petersburg Nuclear Physics Institute, Gatchina 188300, Russia}

\author{V.A. NIKONOV}
\address{National Research Centre "Kurchatov Institute",
Petersburg Nuclear Physics Institute, Gatchina 188300, Russia}

\author{J. NYIRI}

\address{Institute for Particle and Nuclear Physics, Wigner
RCP, Budapest 1121, Hungary }

\maketitle


\begin{abstract}
The LHC energies are those at which the asymptotic regime in
hadron-hadron diffractive collisions ($pp,\pi p,\pi\pi$) might be switched
on. Based on results of the Dakhno-Nikonov eikonal model which is
a generalization of the Good-Walker eikonal approach for a continuous
set of channels, we present a picture for transformation of the
constituent quark mode to the black disk one. In the black disk mode
($\sqrt s \geq 10$ TeV) we have a growth of the logarithm squared type
for  total and elastic cross sections, $\sigma_{tot}\sim\ln^2s$ and
$\sigma_{el}\sim\ln^2s$, and $(\tau={\bf
q}_\perp^2\sigma_{tot})$-scaling for diffractive scattering and
diffractive dissociation of hadrons. The diffractive dissociation
cross section grows as $\sigma_{D}\sim\ln{s}$,
$\sigma_{DD}\sim\ln{s}$, and their relative contribution tends to zero:
$\sigma_{D}/\sigma_{tot}\to 0$, $\sigma_{DD}/\sigma_{tot}\to 0$.
Asymptotic characteristics of diffractive and total cross sections are
universal, and this results in the asymptotical equality of cross
sections for all types of hadrons (the Gribov's universality).
The energy scale for switching on the asymptotic mode is estimated for
different processes.
 \end{abstract}
 \ccode{PACS numbers: 13.85.-t 13.75.Cs 14.20.Dh}


The observation of the total cross section growth at pre-LHC
energies, $\sqrt s \sim 0.2-1.8$ TeV \cite{pre1,pre2,pre3,pre4,pre5},
initiates studies of corresponding models such as that with an increase
allowed by the Froissart bound \cite{Froi} or exceeding it, for
example, by the power-$s$ behavior \cite{Kaid,Land}. The necessity of
the $s$-channel unitarization of scattering amplitudes actualizes the
use of the Glauber approach
\cite{_4glauber1,_4glauber2}. On account of the $s$-channel
rescatterings, the power-$s$ growth of amplitudes is dampened to the
$(\ln^2 s)$-type, see \refcite{Gaisser,Block,Fletcher}. Still, let us
emphasize, exceeding the Froissart bound does not violate general
constraints for the scattering amplitude \cite{azimov}.

Recent measurements at LHC (ATLAS
 \cite{atlas}, CMS \cite{cms}, TOTEM \cite{totem} collaborations) and
cosmic ray data \cite{auger} initiate further interest to $s$-channel
unitarized amplitudes, see, for example,
\refcite{sch-rys,block,ryskin,dremin1,martynov,Ryskin:2009tj,KMR-s3,KMRLHC,Khoze:2013jsa,Maor,Gotsman:2013lya,Ost}.

\section{Diffractive Hadron-Hadron Scattering
in the Dakhno-Nikonov Eikonal Model}

A model for high-energy $\pi p$ and $p^\pm p$ collisions with the
eikonal unitarization of scattering amplitudes was
suggested by Dakhno and Nikonov \cite{DN} and successfully used for
the description of the pre-LHC data. The model takes into account
the quark structure of colliding hadrons, the gluon origin of the
input pomeron and the colour screening effects in collisions. The
model can be considered as a version of the Good-Walker eikonal
approach \cite{GW} for a continuous set of channels.

The extension of the Dakhno-Nikonov model to LHC energies was
carried out in \refcite{ann1} for diffractive $pp$-scattering. The
$pp$ data were re-fitted taking into account new results in the
TeV-region \cite{totem,auger}. The region 5-50 TeV turns out to be
that where the asymptotics starts. It means that the asymptotic
regime should reveal itself definitely at $10^2-10^4$ TeV; an
analysis of the diffractive cross sections at ultrahigh energies was
carried out in \refcite{ann2}.
The total and elastic hadron-hadron cross sections are
approaching the asymptotic values
($\sigma_{tot}^{\asymp}(s)$ and $\sigma_{el}^{\asymp}(s)$)
from bottom to top:
\bea
&& \sigma_{tot}(s)\sim \ln^2s,\quad \sigma_{el}(s)\sim\ln^2s\,, \nn
\\
&& \bigg[\frac{\sigma_{el}(s)}{\sigma_{tot}(s)}
\bigg]_{\ln s\to\infty}\to \frac 12,\qquad
\frac{\sigma_{tot}(s)}{\sigma_{tot}^{\asymp}(s)}< 1,  \qquad
\frac{\sigma_{el}(s)}{\sigma_{el}^{\asymp}(s)}< 1,
\eea
 that gives the illusion of exceeding the Froissart bound (see, for
example, the discussion in \refcite{block3,1212.5096}).

Further, the model predicts that differential elastic cross sections
depend asymptotically on transverse momenta with a relation for
$\tau$-scaling:
 \bea \label{3}
 &&
\frac{1}{ \sigma_{tot}(s)}\, \frac{d\sigma_{el}(\tau)}{d\tau}=
  D(\tau ),
  \qquad
  \int\limits_0^\infty d\tau D(\tau)=
  \frac{\sigma_{el}(s)}{ \sigma_{tot}(s)}\,, \nn\\
  &&
  \tau={\bf q}_\perp^2\sigma_{tot}\propto {\bf q}_\perp^2\ln^2s\,,
\eea
where $\tau $ and $ D(\tau)$ are dimensionless variables.
In $pp$ collision the $\tau$-scaling
is working for $1/\sigma_{tot}\times d\sigma_{el}/d\tau$ within
10$\%$ accuracy for $0< \tau< 4$ at  $\sqrt s \simeq 10$ TeV, and for
$\sqrt s \simeq 100$ TeV this region is extended to $0< \tau<10$.

The model points to the universal behaviour of all total
cross sections (the Gribov  universality \cite{gribov-tot}). It is
the consequence of the universality of the colliding disk structure
(or the structure of parton clouds) at ultrahigh energies. The
question is at what energy range the asymptotic modes are switched
on for different processes.

\begin{figure}
\vspace{1cm}
\centerline{\epsfig{file=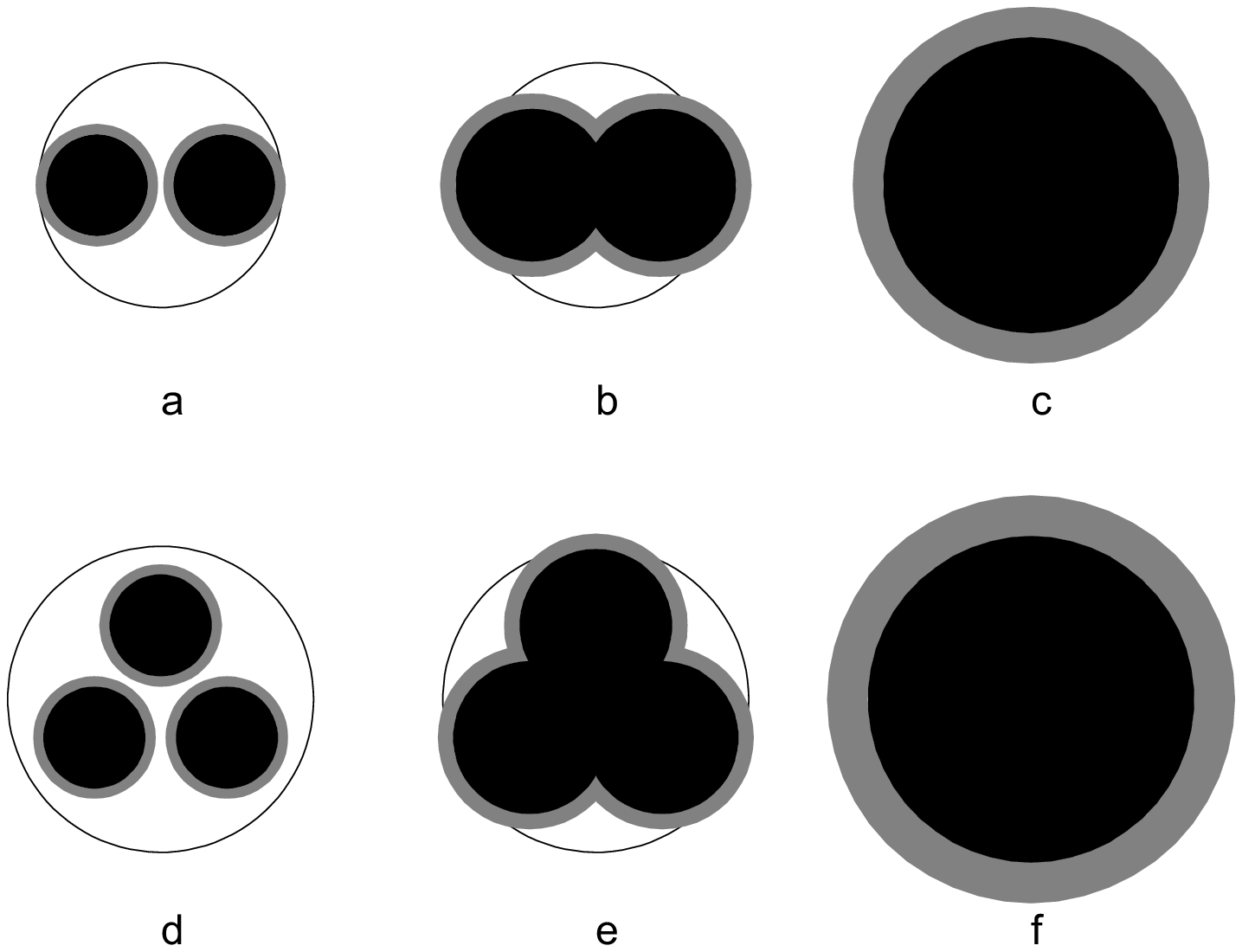,width=0.9\textwidth,clip=on}}
\caption{Meson and proton pictures in the impact parameter space at
 moderately
high energies (a,d) and their transformations with increasing energy
(b,e) to ultrahigh region where universal black disks are created
 (c,f).
\label{disks} }
\end{figure}

Slow switching on of the asymptotic regime in the hadron
diffractive collisions is related both to
the twofold structure of hadrons
 (hadrons are built by constituent quarks and the latter are
formed by clouds of partons) and
the gluon origin of the pomeron.
 In the impact parameter space the
parton clouds fill first the intrinsic hadron domain, see
Figs.~\ref{disks}a,d for pion and proton. At ultra-high energies the situation
is transformed to a one-disk picture, Figs.~\ref{disks}c,f that
manifests itself in two-step asymptotics. The energy of this
transformation is that of LHC, Figs.~\ref{disks}b,e.

The calculations we have carried out demonstrate a comparatively
fast approach of $\sigma_{tot}(s)$ to the asymptotic behavior being
in contrast to $\sigma_{el}(s)$.
The slow increase of
$\sigma_{el}(s)$ means a slow approach of
$\sigma_{inel}(s)=\sigma_{tot}(s)-\sigma_{el}(s)$ to the asymptotic
mode.

A fast approach to the asymptotic
mode is observed for the sum of elastic and quasi-elastic
cross sections
(elastic, single diffractive dissociations and double diffractive
dissociation). That emphasizes the importance of the study of
inelastic diffractive processes.
We demonstrate that the diffractive dissociation cross sections are
increasing at asymptotic energies as
($\sigma_{D}\propto\ln{s}$, $\sigma_{DD}\propto\ln{s}$)
while their relative contribution tends to zero
($\sigma_{D}/\sigma_{tot}\to 0$, $\sigma_{DD}/\sigma_{tot}\to 0$).
Specifically, the cross section for the diffractive production of
$N_{\frac 12^+}(1440)$ is estimated
 as
$\left(\frac{1}{10}\div\frac 12\right)
\cdot 0.6\ln{\frac{s}{1.2 {\rm GeV}^2}}$ mb
if this state is a radial excitation of a nucleon \cite{ann2}.

The effect for inclusive cross sections due to the change of the
regime, from the constituent quark collision picture to that with a
united single disk, was discussed in
\refcite{ani-she,ani-lev-rys,ani-bra-sha} when definite indications
about hadron cross section growth appeared. It was emphasized that
the approach to the single black disk regime should change
probabilities of the production of hadrons in the fragmentation region
(hadrons with $x=p/p_{in}\sim 1$), for more details see
also ref. \refcite{book2}.

The increase of the black disk radius, $R_{black}\propto 2\sqrt{
\Delta\alpha'_P}\ln s$, is determined by parameters of the leading
$t$-channel singularity, that are the pomeron intercept
$\alpha(0)=1+\Delta$ (with $\Delta>0$) and the pomeron trajectory
slope $\alpha'_P$. The $s$-channel unitarization of the scattering
amplitude damps the strong pomeron pole singularity transforming it
into a multipomeron one. Therefore, we face the intersection of
problems of the gluon content of the $t$-channel states at ultrahigh
energies and the physics of glueball states - at present the
glueball states are subjects of intensive investigations, see, for
example \refcite{vva-usp,book3,klempt-zaitsev,ochs} and references
therein. Studies of phenomenons related to glueballs and multigluon
states at small $|t|$ (or at small masses) are enlightening for
understanding the confinement singularity - see the discussion in ref.
\refcite{conf-PR}. The large value of the mass of the soft effective
gluon (and the corresponding value of the low-lying glueballs) and the
slow rate of the black disk increase appear to be related phenomena.

The model is based on the hypothesis of the gluon origin of the
$t$-channel forces, and these gluons form pomerons. Mesons
(two-quark composite systems) and baryons (three-quark composite
systems) scatter on the pomeron cloud. It is supposed that the
pomeron cloud is materialized as a low-density gas, and
pomeron-pomeron interactions, as well as $t$-channel transitions
$P\to PP$, $P\to PPP$ and so on, can be neglected.

\begin{figure}
\centerline{\epsfig{file=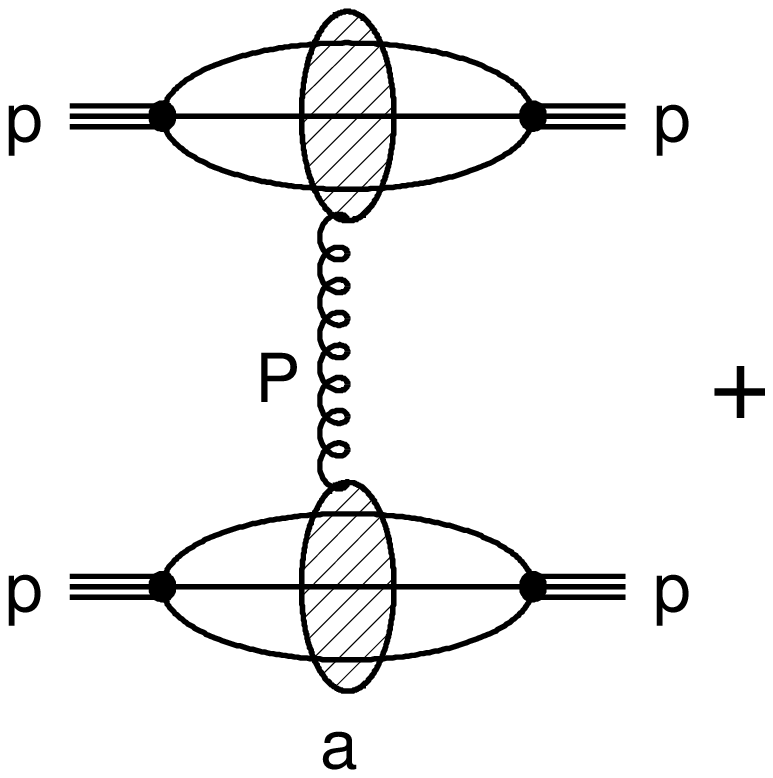,width=3.5cm}\hspace{0.2cm}
            \epsfig{file=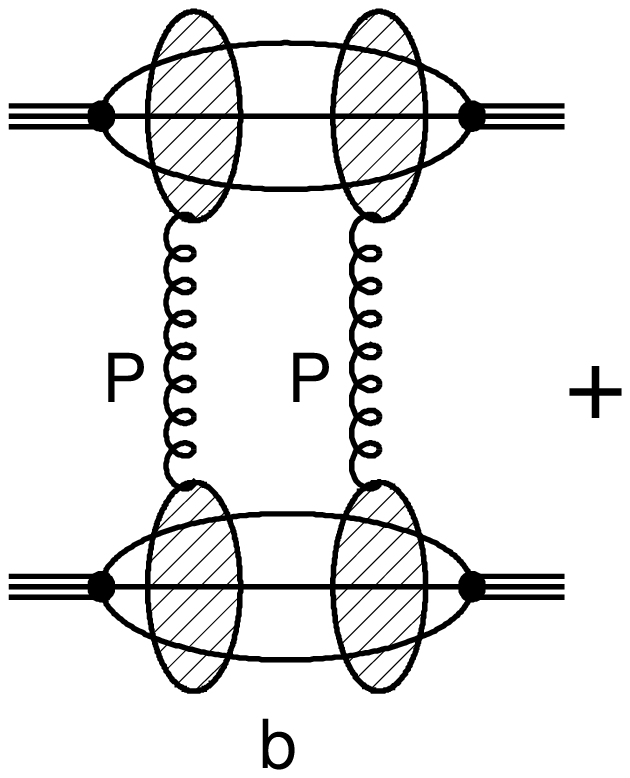,width=3.5cm}\hspace{0.2cm}
            \epsfig{file=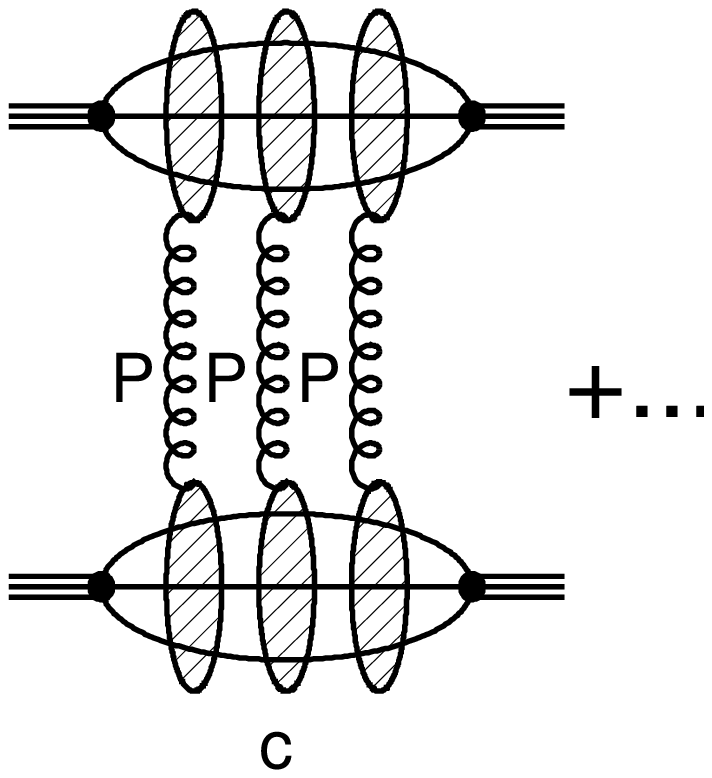,width=3.5cm}}
\vspace{0.5cm}
\centerline{ \epsfig{file=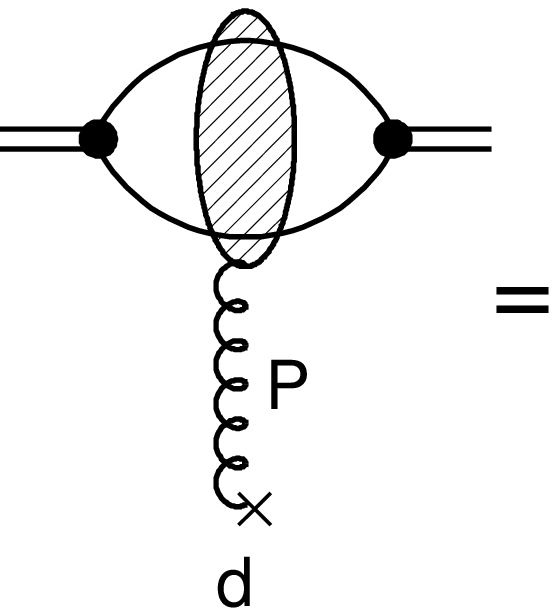,height=3.3cm}
             \epsfig{file=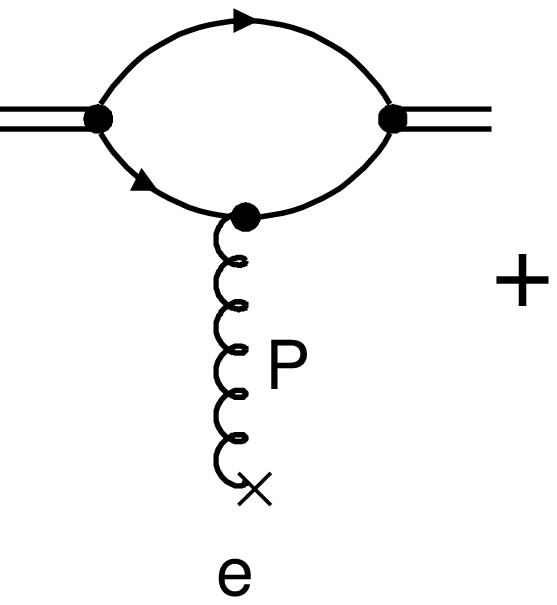,height=3.3cm}
             \epsfig{file=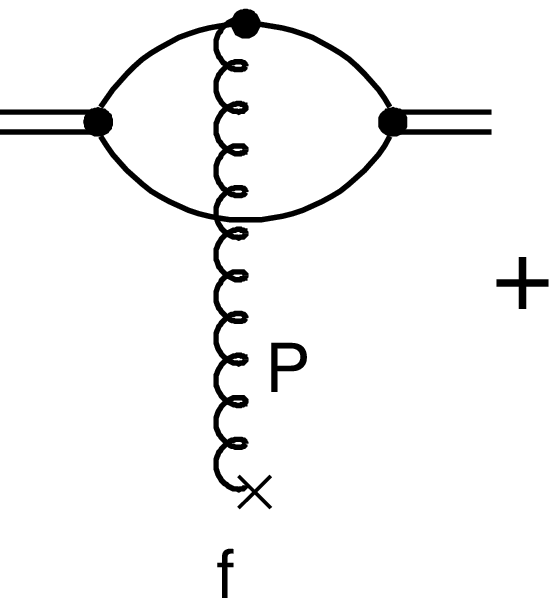,height=3.3cm}
             \epsfig{file=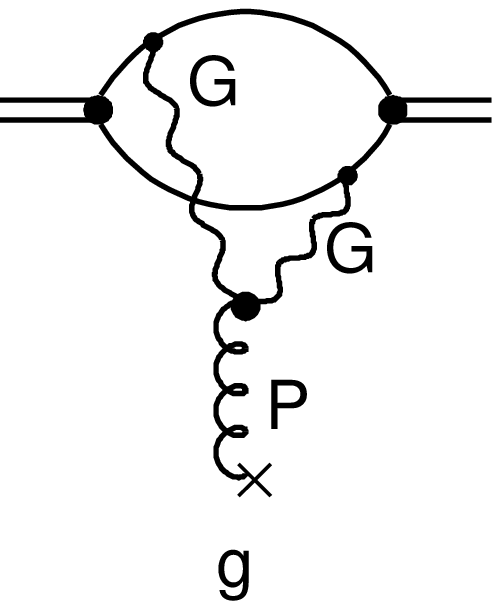,height=3.3cm}}
\vspace{0.5cm}
\centerline{\epsfig{file=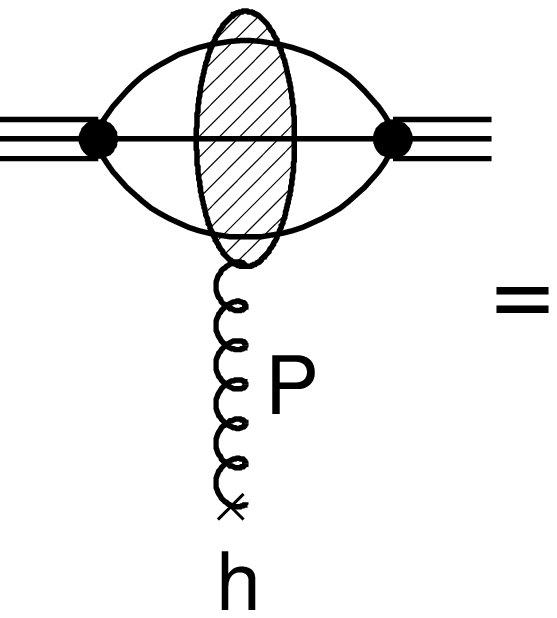,height=2.5cm}
            \epsfig{file=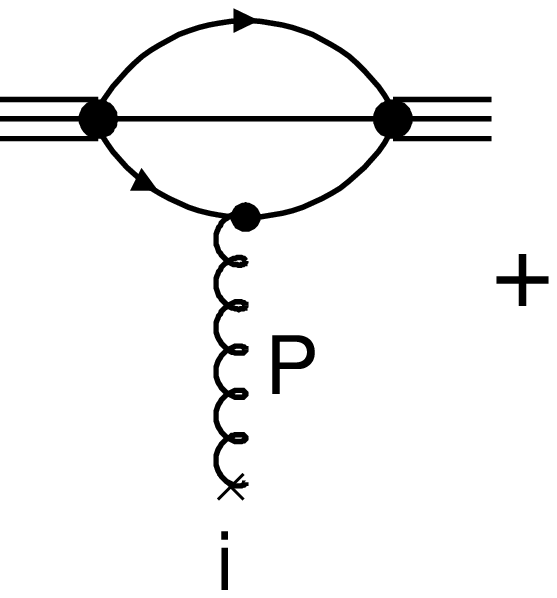,height=2.5cm}
            \epsfig{file=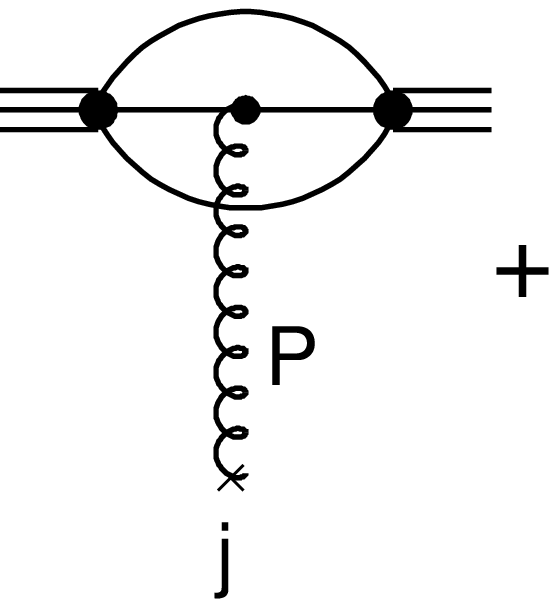,height=2.5cm}
            \epsfig{file=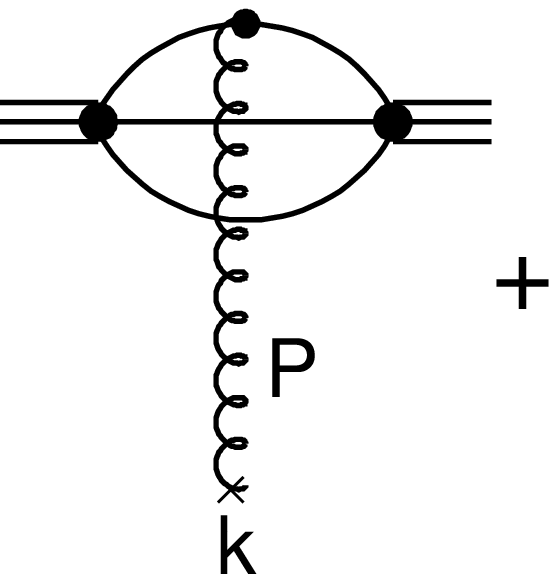,height=2.5cm}
            \epsfig{file=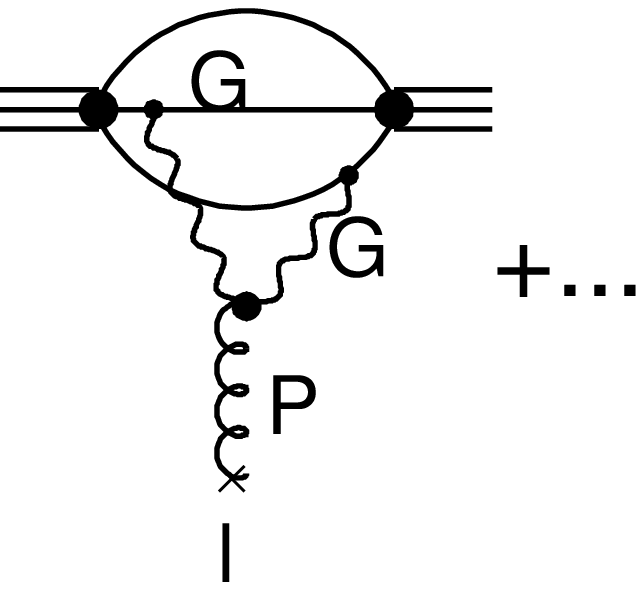,height=2.5cm}}
\caption{Hadron-hardron collisions: (a)--(c) diagrammatic
representation of the $\pi p$ scattering
amplitude as a set of $s$-channel pomeron ($P$) interactions;
(d)--(g) pion-pomeron vertex and its deciphering:
pomeron-quark terms and  $GGP$ term which describes interaction of
gluons ($G$) of pomeron with both quarks of pion;
(h)-(l) proton-pomeron vertex and corresponding $P$ and $GGP$
exchanges.
\label{f2} }
\end{figure}

Consequently, the
hadron-hadron scattering amplitude is determined by the set of diagrams
with multiple pomeron exchanges (for $\pi p$ scattering it is shown in
Fig.~\ref{f2}a,b,c). Pion-pomeron and proton-pomeron coupling
diagrams are shown in Fig.~\ref{f2}d-g and  Fig.~\ref{f2}h-l.
Gluon structure
of pomeron provides the colour screening effect in couplings
\cite{adnPR}.

The convergence of the coupling diagrams is guaranteed by
vertices $\pi\to q\bar q$ and $p\to qqq$, or quark wave functions
of hadrons $\varphi_{\pi}^2({\bf r}_1,{\bf r}_2)$ and
$\varphi_{p}^2({\bf r}_1,{\bf r}_2,{\bf r}_3)$,
where ${\bf r}_a$ are the transverse coordinates of quarks.
Quark wave functions
can be determined using the corresponding form factors,
an example for such a determination is given in \refcite{AMN}.

The shape of the black disk is formed by the $t$-channel pomeron
exchanges: in \refcite{DN,ann1}  it is a two-pole presentation of
the QCD-motivated pomeron with intercepts
$\alpha(0)=1$ and $\alpha(0)=1+\Delta$. The
two-pole pomeron
exchange is popular from the sixties till now, see for example ref.
\refcite{DL}.
 In the ultra-high energy region
the leading pomeron dominates,
according to fit \cite{ann1} its trajectory reads:
\bea
&&
\alpha_P({\bf q}^2_\perp)\simeq 1+\Delta-\alpha'_P{\bf q}^2_\perp \\
&&
\Delta=  0.273\pm 0.011,\quad
\alpha'_P= 0.129 \pm 0.007
    ({\rm GeV/c})^{-2}    \nn
\eea

Meson-pomeron and baryon-pomeron couplings vanish for squeezed
configurations of quarks: for mesons at
$|{\bf r}_{12}|\to 0\, $ and for baryons
at $|{\bf r}_{12}|\to 0,\, |{\bf r}_{13}|\to 0$ where
$|{\bf r}_{ij}|= | {\bf r}_{i}- {\bf r}_{j}|$ that is a result of
interplay of one-quark interaction couplings and two-quark
interaction \cite{adnPR}. Corresponding diagrams which guarantee
the colour screening are shown in
Figs.~\ref{f2}d-g for mesons and
Figs.~\ref{f2}h-l for baryons.

\begin{figure}[h]
\centerline{\epsfig{file=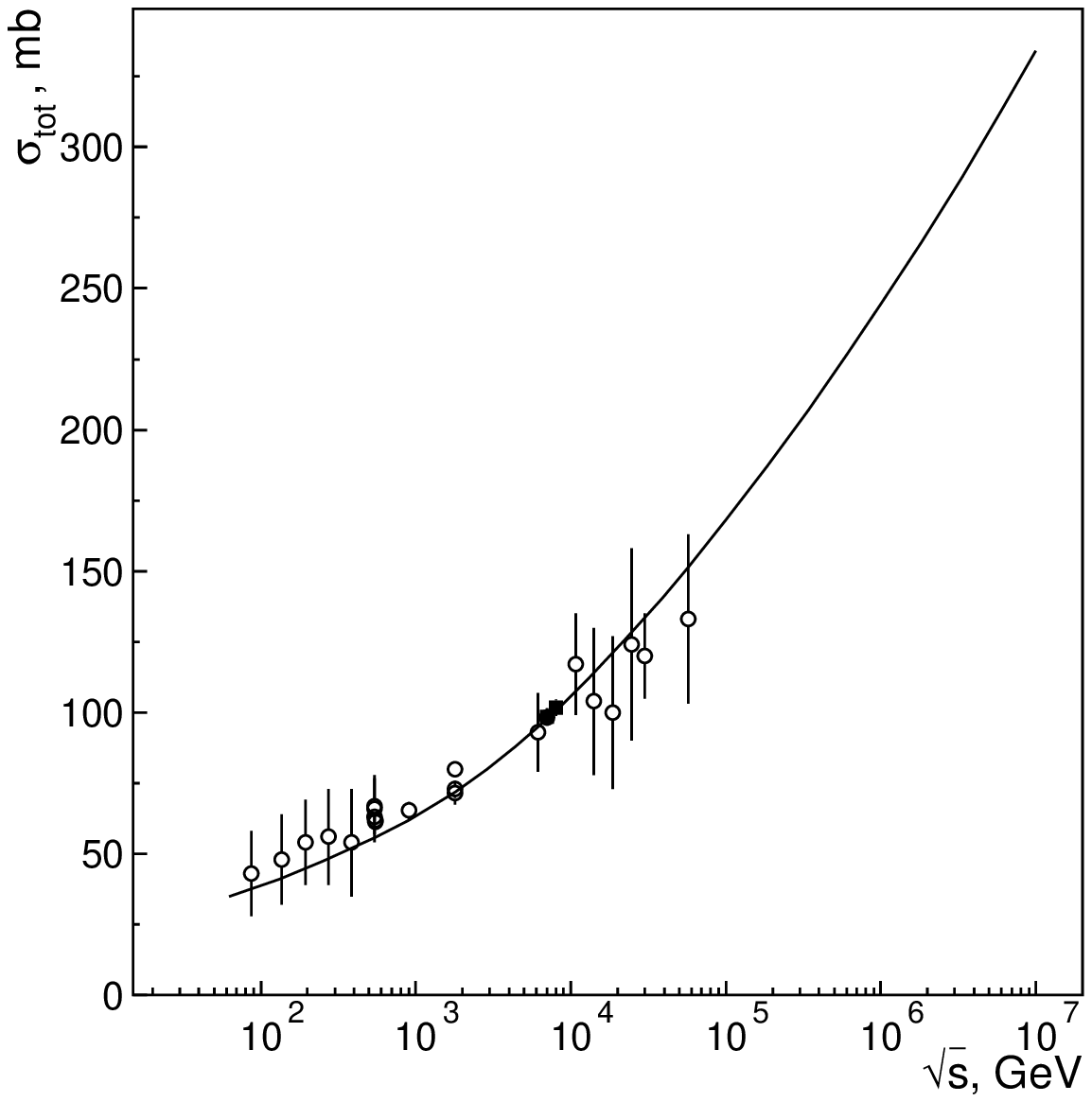,width=0.48\textwidth}
            \epsfig{file=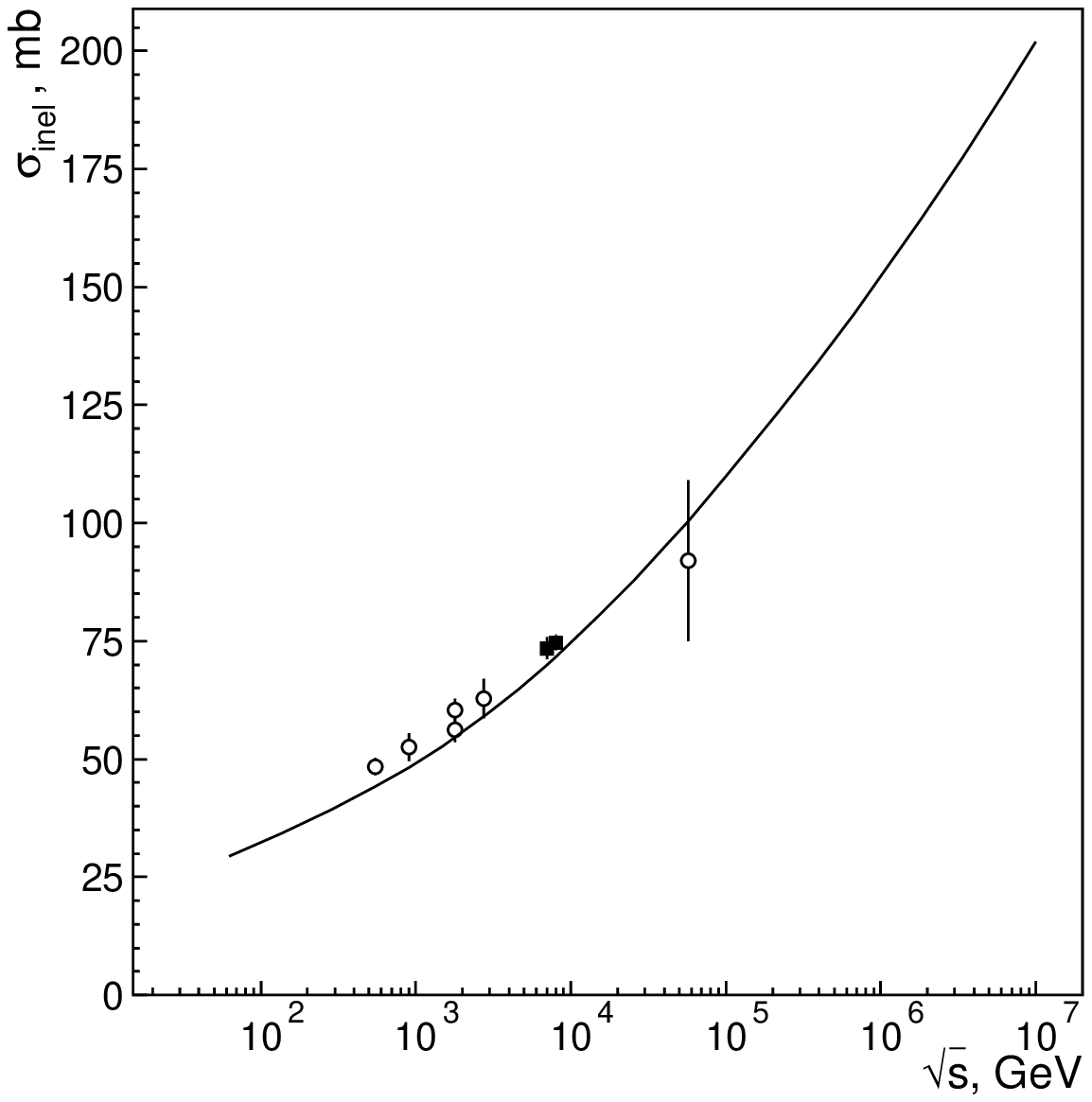,width=0.48\textwidth}}
\caption{Total and inelastic cross section data
\cite{pre1,pre2,pre3,pre4,pre5,totem,auger}
and fit in the Dakhno-Nikonov model.} \label{f3} \end{figure}

\begin{figure}[h]
\centerline{\epsfig{file=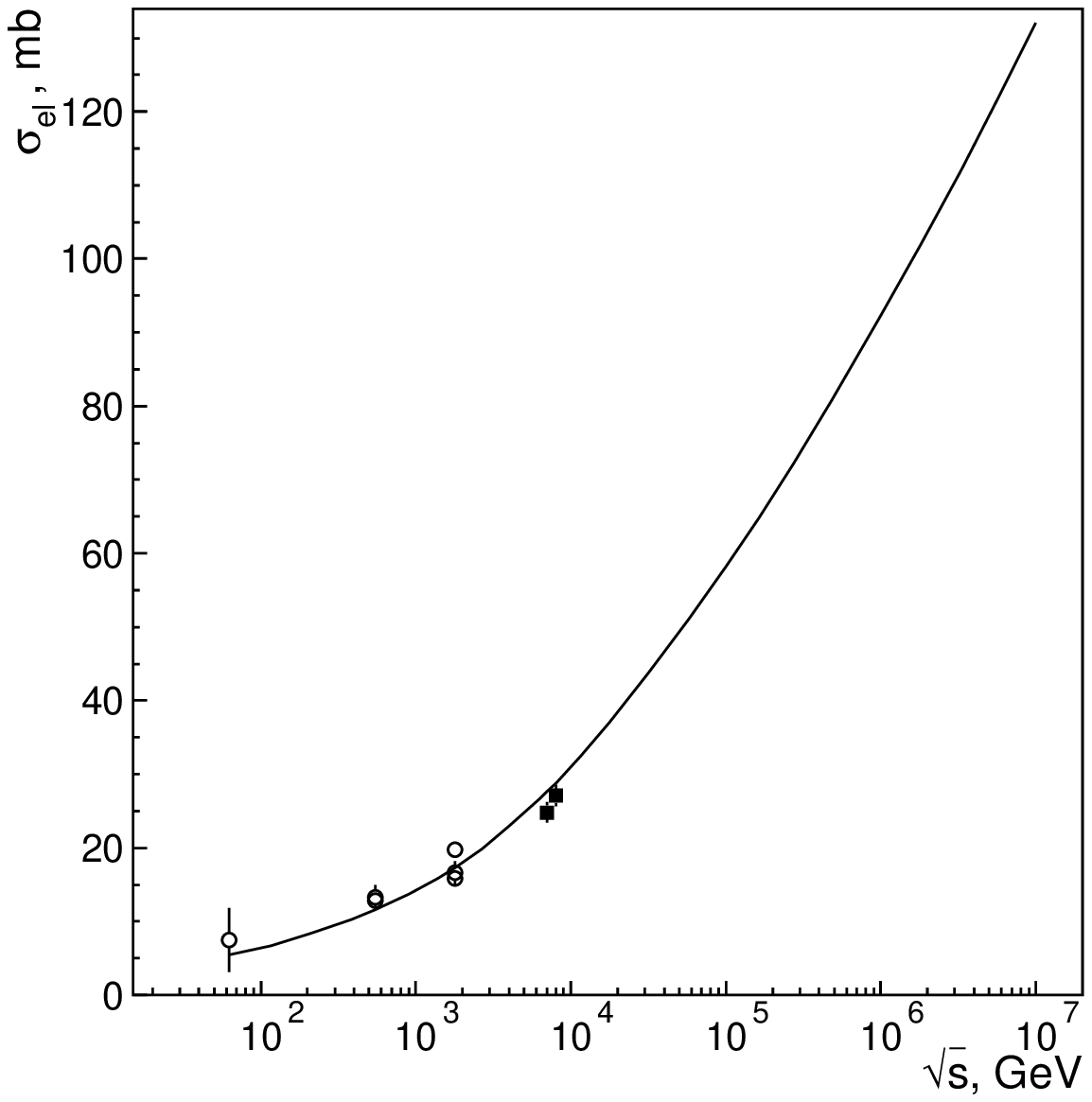,width=0.48\textwidth}
            \epsfig{file=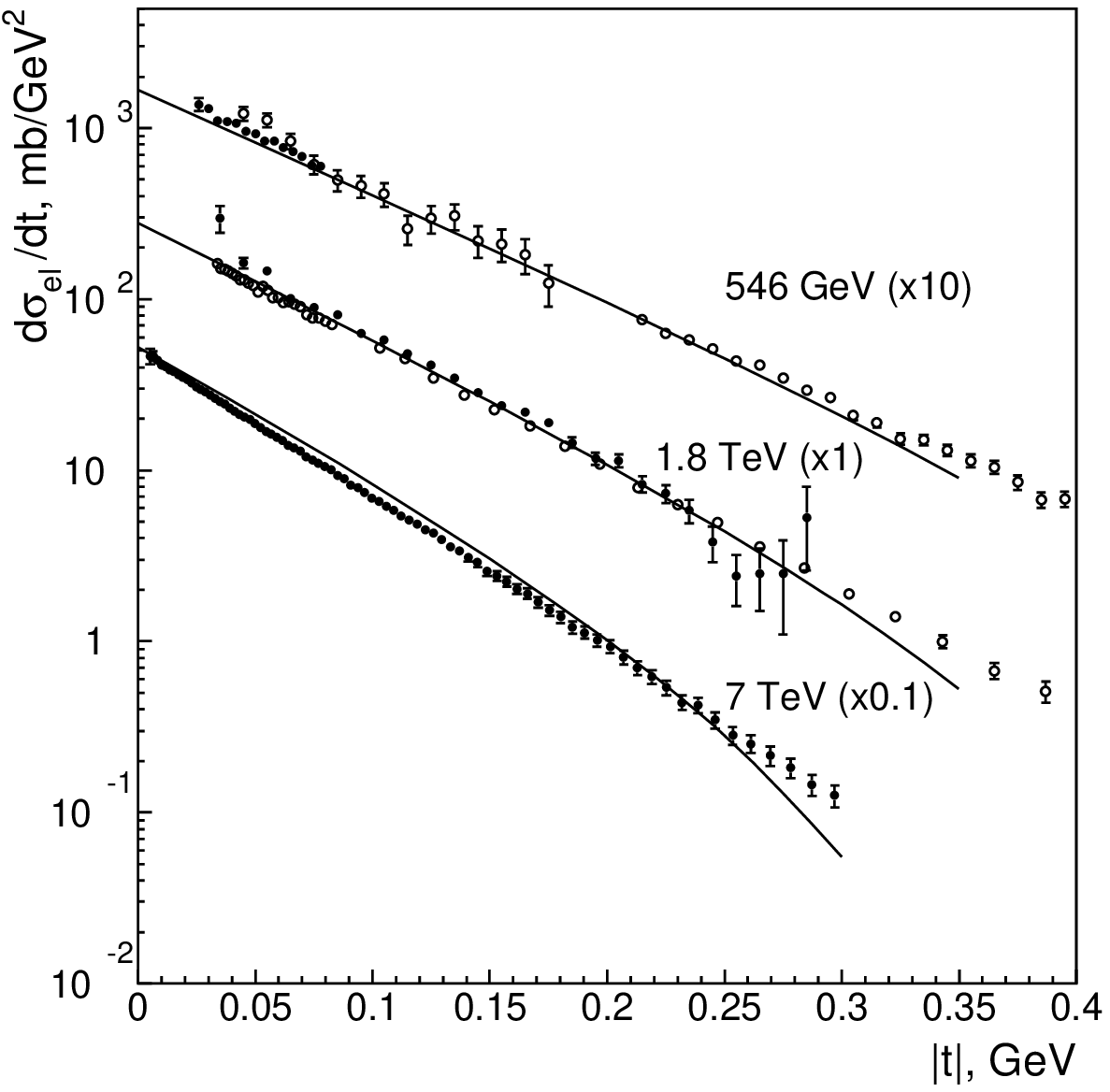,width=0.48\textwidth}}
\caption{Elastic cross section and
differential cross sections
$ d\sigma_{el}/d{\bf q}^2_\perp$
  at $\sqrt s =0.546,\, 1.8,\, 7.0$
 TeV and their descriptions in the fit.
\label{f4} }
\end{figure}

\begin{figure}[h]
\centerline{
 \epsfig{file=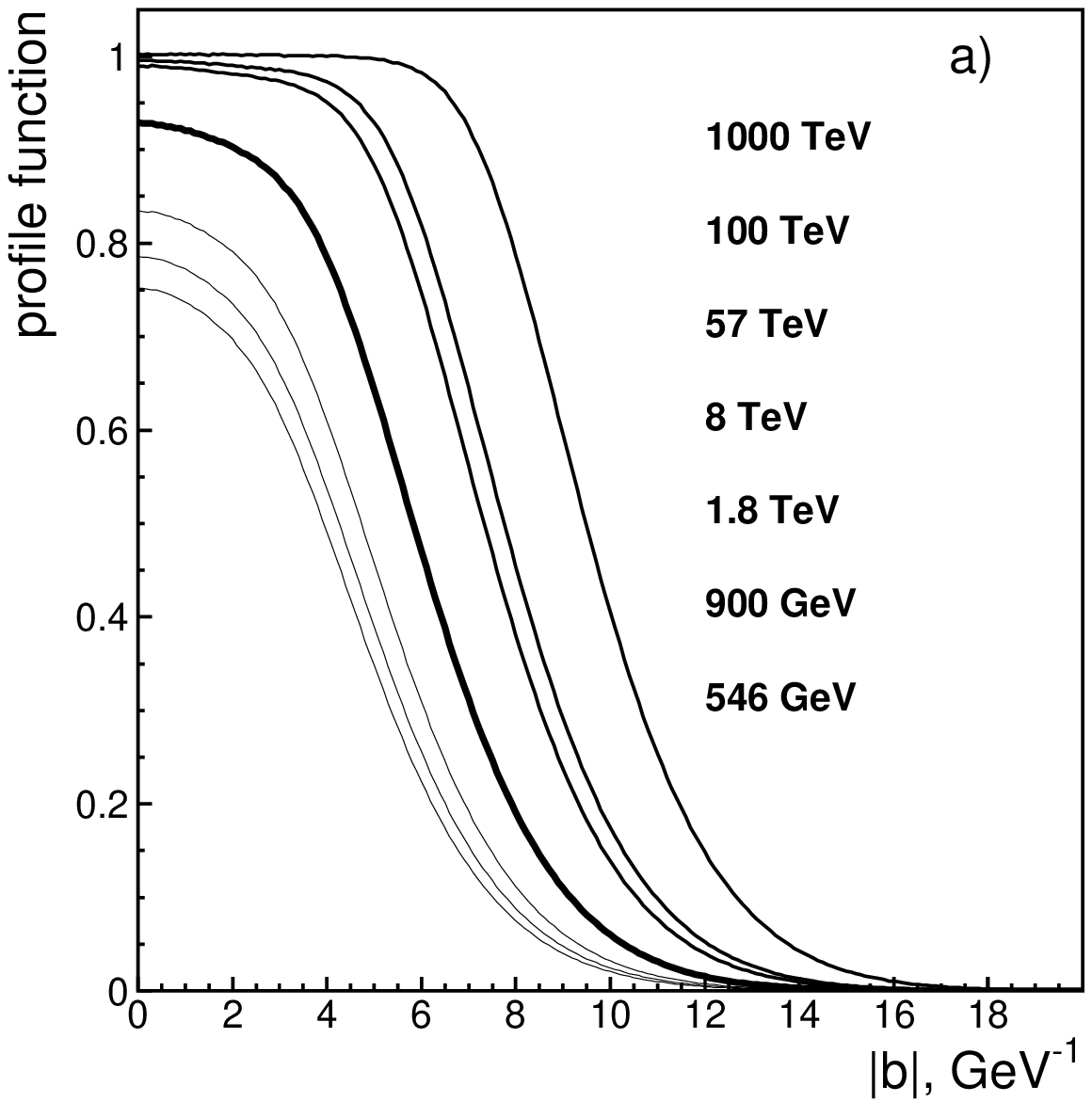,width=0.48\textwidth}
\hspace{0.3cm}
     \epsfig{file=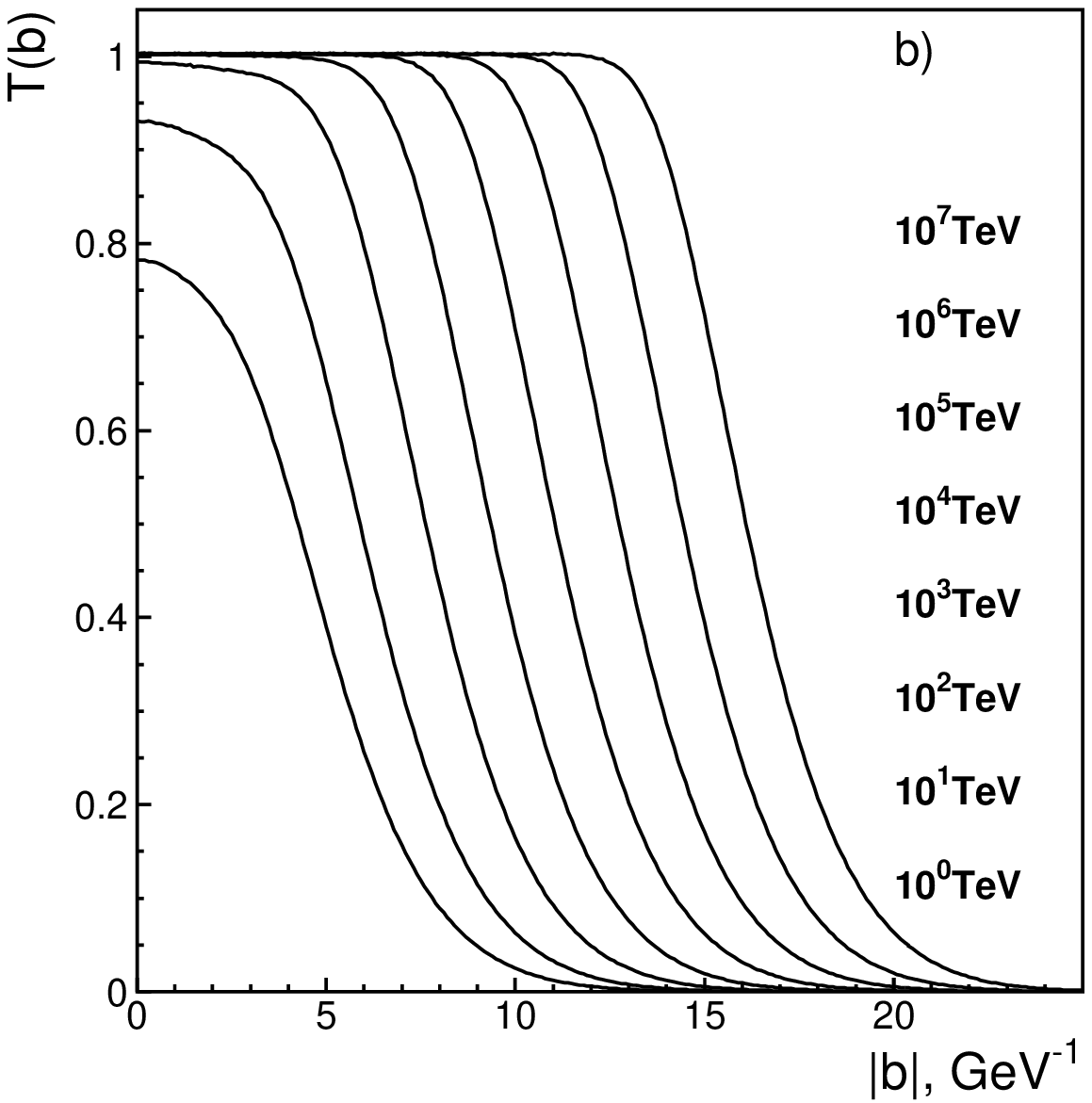,width=0.48\textwidth}      }
\caption{a) Profile functions $T(b)$ determined in Eq.
(\ref{p9})
 at preLHC (0.546-1.8 TeV), LHC (8.0 TeV),
Pierre Auger Collaboration (57 TeV) and ultra-high
(100-1000 TeV) energies.
b)Profile functions $T(b)$ at a set of energies $\sqrt{s}=1,10,100,...,10^7$ TeV.}
\label{f5}
\end{figure}

Results of description of ultrahigh energy data, that are $pp$ data
including those at LHC energies \cite{totem} and cosmic ray ones
\cite{auger}, are demonstrated in Figs.~\ref{f3},\ref{f4}.

In Fig.~\ref{f3} we show total and
inelastic cross sections, the fit gives a good approximation to data
at $\sqrt{s}\sim 50-5\cdot 10^4$ GeV; we also show predictions
 for the region $\sqrt{s}\la 10^5$ GeV.
The fit gives reasonably good level of description of $\sigma_{el}$
and the differential cross section $d\sigma_{el}/d{\bf q}^2_\perp$,
see Fig.~\ref{f4}.

The fit curves for $\sigma_{tot}$, $\sigma_{inel}$ and $\sigma_{el}$
look like as the LHC energy is a turning
point for the cross section growth regime. Definitely it is seen
when considering the scattering amplitude in the impact
parameter space.

\section{Scattering Amplitude in the Impact Parameter Space}

In Fig.~\ref{f5} we show
profile functions $T(b)$ determined as
\bea \label{p9}
&&
\sigma_{tot}=2\int d^2b\; T(b)=2\int d^2b
\Big[1-e^{-\frac12\chi(b)}\Big],\\
&&
4\pi\frac{d\sigma_{el}}{d{\bf q}^2_\perp}=
A^2({\bf q}^2_\perp),\quad
A({\bf q}_\perp)=\int d^2b e^{i{\bf b}{\bf q}_\perp} T(b)
\nn
\eea
The profile functions $T(b)$ at energies $0.5 - 10^3$ TeV
are shown in Fig.~\ref{f5}a -
it is seen that the profile function saturation
mode, $T(b)\to 1$, works at ultra-high energies ($100-1000$ TeV), at
LHC energies the saturation mode is only starting.

The profile functions at $0.546-1.8$ TeV  illustrate the two-fold
structure of proton, corresponding to Fig.~\ref{disks}d. At that
energies the absorption area is dominantly inside the proper nucleon
size, $b\leq 4$ GeV$^{-1}$, and only the blackness level is increasing
with energy.

\begin{figure*}[ht]
\centerline{\epsfig{file=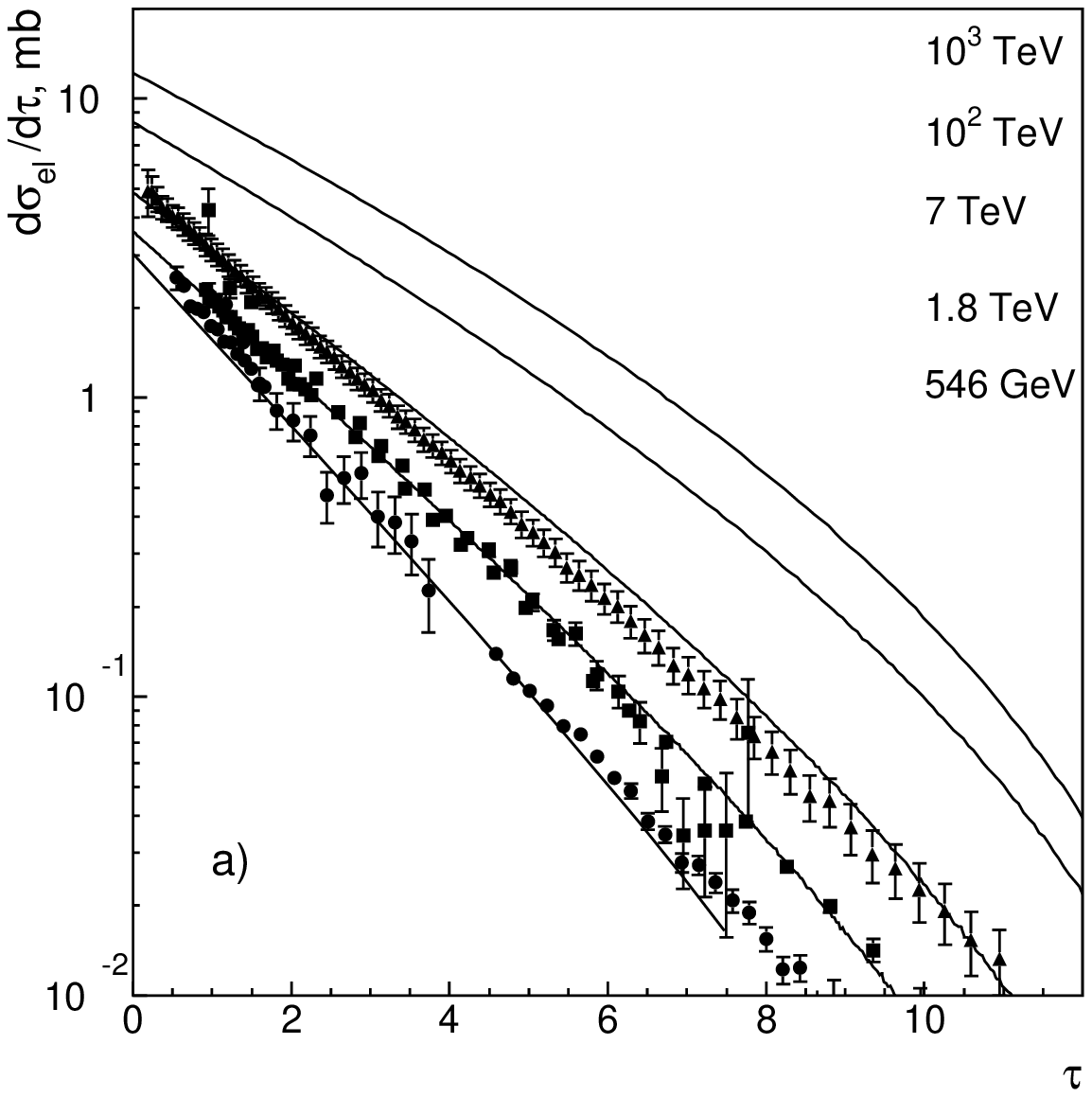,width=0.48\textwidth}
           \epsfig{file=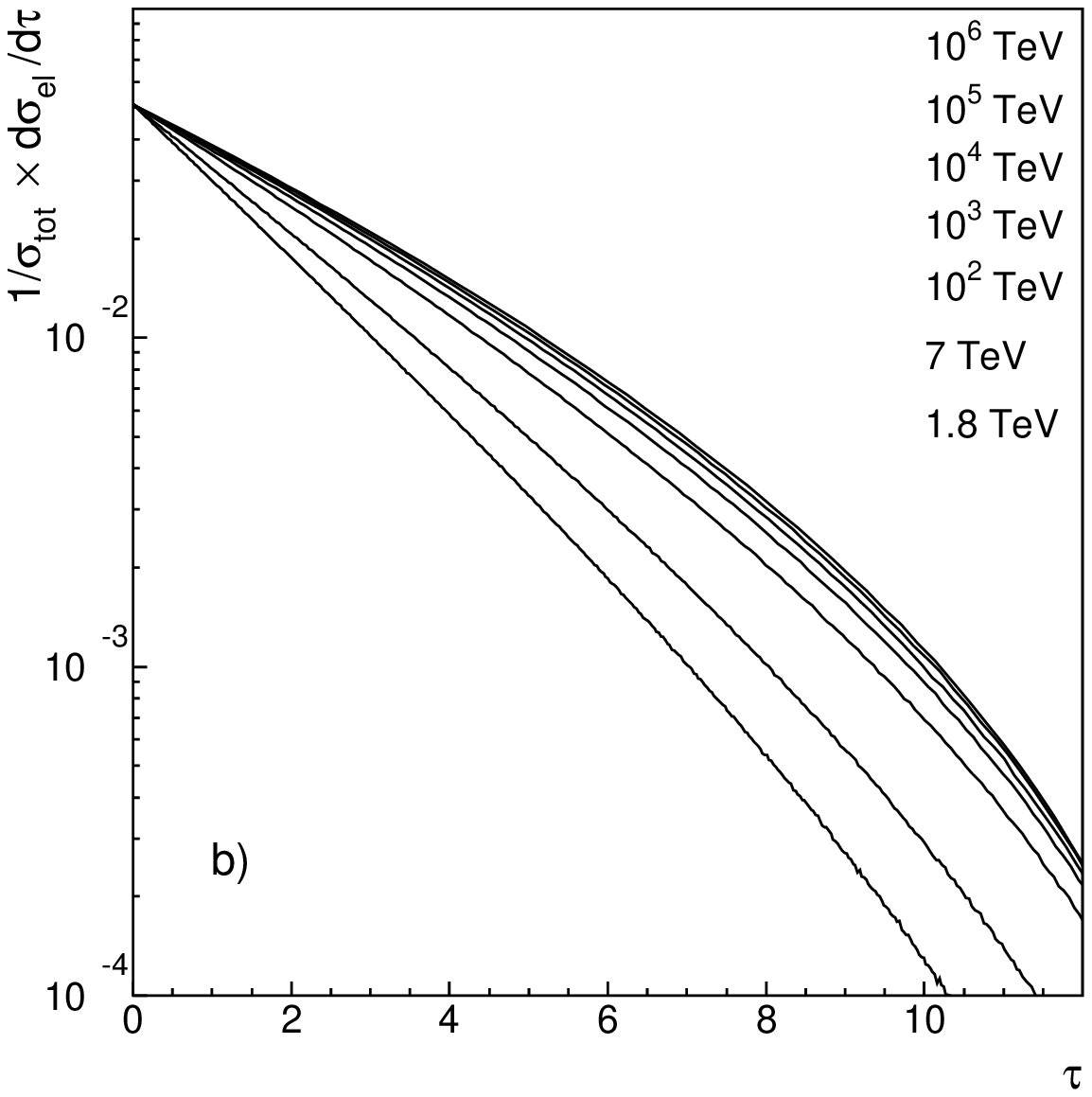,width=0.48\textwidth}            }
\caption{ a) Differential cross sections
$ d\sigma_{el}/d\tau$, where $\tau=\sigma_{tot}{\bf q}^2_\perp$,
  at $\sqrt s =0.546,\, 1.8,\, 7.0$
 TeV and their descriptions in the Dakhno-Nikonov model;
 b)  Calculated differential cross sections
$1/\sigma_{tot}\times d\sigma_{el}/d\tau$ at
$\sqrt{ s} = 1.8,\,7,\, 100,\, 1000,...10^6$ TeV
and their approaching
to the $\tau$-scaling limit.
}
\label{f7}
\end{figure*}

 \section{Diffractive Cross Section and the
 $\tau$-Scaling Limit}

In Fig.~\ref{f7}a we show $d\sigma_{el}/d\tau$ with
$\tau=\sigma_{tot} {\bf q}^2_\perp $ at ISR and LHC energies; the
approach of ${1}/{\sigma_{tot}}\times{d\sigma_{el}}/{d\tau}$ to the
$\tau$-scaling limit is demonstrated in Fig.~\ref{f7}b. The
$\tau$-scaling, being inherent with the Dakhno-Nikonov model,
 works with a dimensionless object,
${1}/{\sigma_{tot}}\times{d\sigma_{el}}/{d\tau}$,
in terms of the variable
$\tau$. So the scaling regime works at constrained region of
non-large $\tau$ (or $\tau\le 10$), it is the region of the diffractive cone.

The region of large $\tau$ attracts attention as well, see, for
example, \refcite{azimov,dremin1,uzhi,dremin2}. However, the
large-$\tau$ region
can be formed by next-to-leading $t$-channel singularities which are
related to other physics, not physics of the diffractive processes.

A variable similar to $\tau$, which also can be used for the description
of the discussed diffractive distributions,
$\tilde t={\bf q}^2_\perp \ln^2s$, was used in \refcite{azimov,dremin1}.
In \refcite{dremin1} the diffractive distribution was considered in
terms of such a variable but with a dumping factor at small $\tilde t$
that means strengthening the role of non-peripheral interactions of the
large-$\tau$ region.

 \section{The Black Disk Radius and $\ln^2 s$-Growth
 of Total and Elastic Cross Sections}

\begin{figure}[h]
\centerline{\epsfig{file=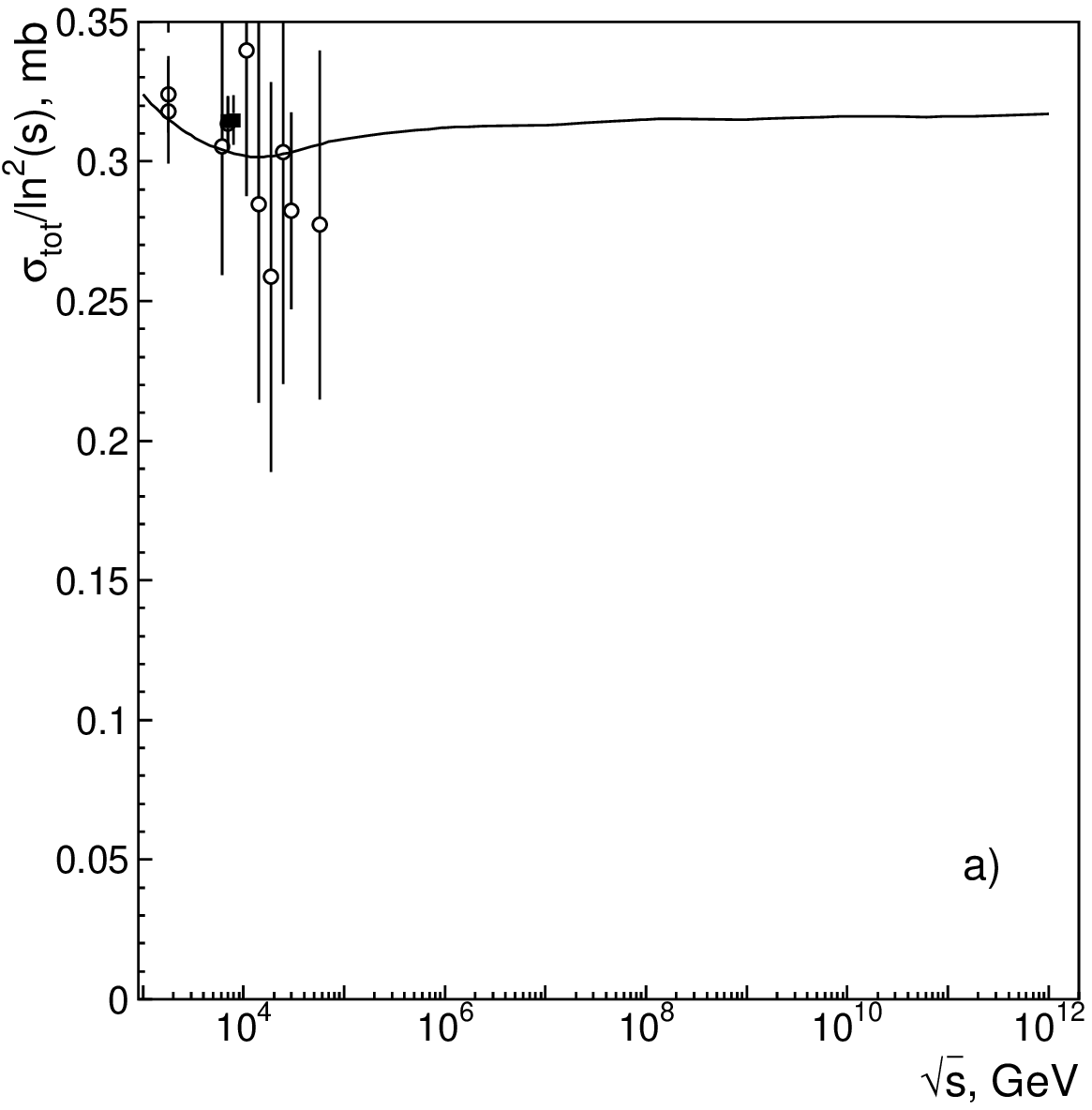,width=0.48\textwidth}
            \epsfig{file=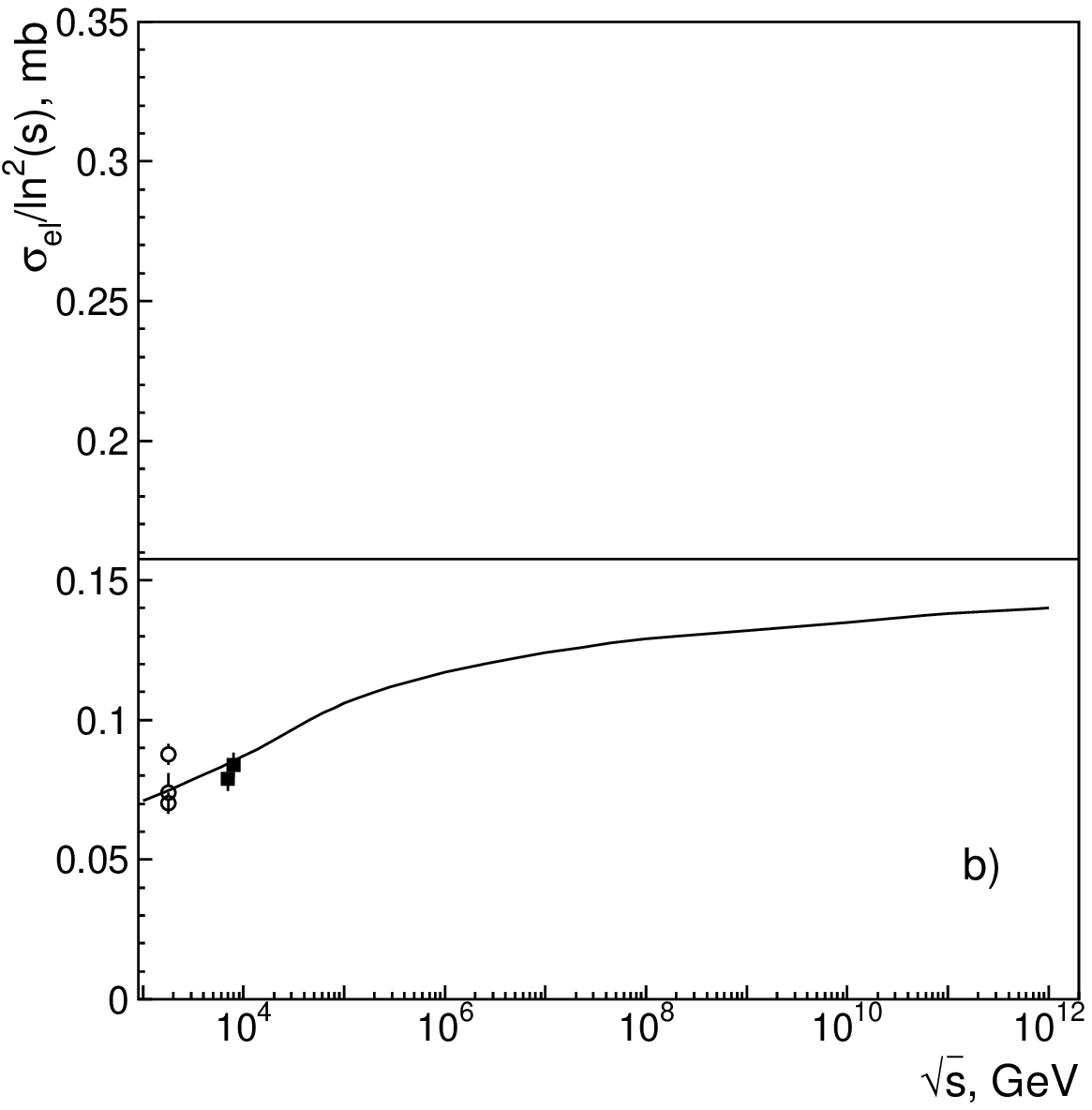,width=0.48\textwidth}}
\caption{
Values for $\sigma_{tot}(pp)/\ln^2s/s_0$ and
$\sigma_{el}(pp)/\ln^2s/s_0$ (where $\sqrt s_0= 1$ GeV) in the
Dakhno-Nikonov model versus data
\cite{pre1,pre2,pre3,pre4,pre5,totem,auger}. The straight line is the
asymptotic limit for the elastic cross section:
$\sigma^{\asymp}_{el}(s)/\ln^2s=\frac12\sigma^{\asymp}_{tot}/\ln^2s$.
}
\label{f8}
\end{figure}


 At $\ln s>>1$, when the asymptotic regime works, there are two
  clear regions in the $b$-space (Fig.~\ref{f5}):
 with $T(b)\simeq 1$ (black disk area) and $T(b)\simeq 0$ (transparent
 area).
 Conventionally we determine these areas by the constraints:
$T(b)>0.97$ and $T(b)<0.03$. Then
\bea  \label{yf15}
&&
{\bf b}^2 < 4\Delta\alpha'_P\ln^2{\frac {s}{s_-}}\,,\qquad
{\rm with}\quad T(b)>0.97,
\nn \\
&& {\bf b}^2 > 4\Delta\alpha'_P\ln^2{\frac {s}{s_+}}\,,\qquad {\rm
with}\quad T(b)< 0.03\,,
\quad
\eea
 which gives the black disk radius:
\be \label{yf16}
R_{black}=2\sqrt{ \Delta\alpha'_P}\,\ln{\frac
{s}{s_R}}\,,\qquad \sqrt{s_{R}}\simeq 80\, {\rm GeV}\,,
\ee
with $s_-< s_{R}< s_+$.  The black disk radius depends on parameters of the leading pomeron
only (factor $2\sqrt{\Delta\alpha'_P}\simeq 0.08$ fm) that results in
Gribov's universality of hadronic total cross
sections at asymptotic energies \cite{gribov-tot}.

%

An approach of $\sigma_{tot}(s)$ and $\sigma_{el}(s)$ to asymptotic
values is shown on Fig.~\ref{f8} - the calculations demonstrate a
relatively fast appearance of asymptotics in $\sigma_{tot}(s)$ in
contrast to $\sigma_{el}(s)$. A slow increase of $\sigma_{el}(s)$
means a slow approach of
$\sigma_{inel}(s)=\sigma_{tot}(s)-\sigma_{el}(s)$ to the asymptotic
behaviour.

\section{Diffractive dissociation cross sections}

\begin{figure}[h]
\centerline{\epsfig{file=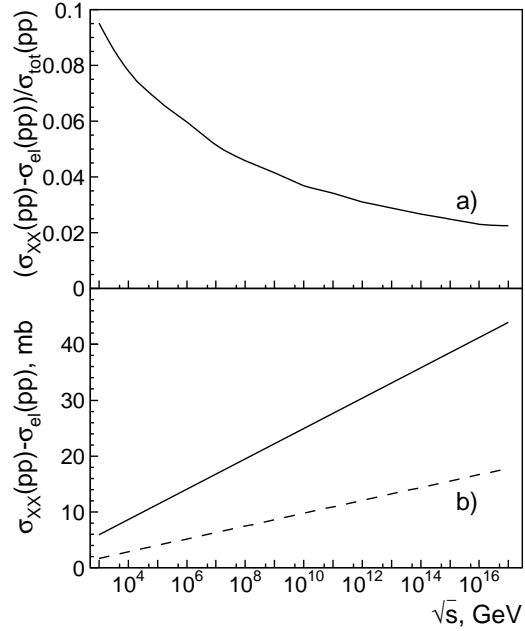,width=7cm}}
\caption{Diffractive processes:
a) the ratio
$\bigg(2\sigma_{D(p)}(pp)+\sigma_{D(p)D(p)}(pp)\bigg)\,/\sigma_{tot}(pp)$
determined accoding to Eq. (\ref{ddpp})
and
b)  differences
$\sigma_{X(p)X(p)}(pp)-\sigma_{el}(pp)=
2\sigma_{D(p)}(pp)+\sigma_{D(p)D(p)}(pp)$ (solid line)
and $\sigma_{X(p)}(pp)-\sigma_{el}(pp)=\sigma_{D(p)}(pp)$ (dashed line),
presented in Eq. (\ref{ddpp8}).
 \label{dpp_02}
 }
 \end{figure}

The relative weight of quasi-elastic processes falls with energy growth
as $1/\ln s$ ; this is in agreement with
$\sigma_{tot}/\sigma_{el}\to 2$.
In Fig.~\ref{dpp_02}a we demonstrate the ratio
\be \label{ddpp}
\frac{\sigma_{X(p)X(p)}(pp)-\sigma_{el}(pp)}{\sigma_{tot}(pp)}
=\frac{2\sigma_{D(p)}(pp)+\sigma_{D(p)D(p)}(pp)}{\sigma_{tot}(pp)}
\sim\frac{1}{\ln s}\, ,
 \ee
it tends to zero at $\ln s\to\infty$.
In Fig.~\ref{dpp_02}b we show the sum of the quasi-elastic cross sections
$2\sigma_{D(p)}(pp)+\sigma_{D(p)D(p)}(pp)$ (solid line) and the single
diffractive dissociation cross section
$\sigma_{D(p)}(pp)$ (dashed line).
Calculations show that the diffractive dissociation cross sections increase as follows:
\bea
&&
2\sigma_{D(p)}(pp)+\sigma_{D(p)D(p)}(pp)=
\sigma_{X(p)X(p)}(pp)-\sigma_{el}(pp)\simeq 0.58
\ln\frac{s}{104\; {\rm GeV }^2}\; {\rm mb},
\nn \\
&&
\sigma_{D(p)}(pp)=
\sigma_{X(p)}(pp)-\sigma_{el}(pp)\simeq 0.25
\ln\frac{s}{1007\; {\rm GeV }^2}\; {\rm mb},
\nn \\
&&
\sigma_{D(p)D(p)}(pp)\simeq ( 0.08\ln s +0.5 )\quad
 {\rm mb}.
  \label{ddpp8}
\eea
$s$ is given in GeV$^2$ units. At 7 TeV calculations give
$2\sigma_{D(p)}(pp)+\sigma_{D(p)D(p)}(pp)\simeq 7.6$ mb
$\sigma_{D(p)}(pp)\simeq 2.7$ mb.
The values tell us that the single diffraction dissociation
gives the main contribution to the quasi-elastic diffractive processes,
while the double diffraction increases more slowly. The corresponding
cross sections are related to semitransparent disk rim:
\bea
&&
2\sigma_{D(p)}(pp)+\sigma_{D(p)D(p)}(pp) =2\pi R_{\rm black}\,\delta R(2D+DD),
\nn \\
&&
\delta R(2D+DD)=0.16\pm 0.04\; {\rm fm},
\nn
\\
&&
\sigma_{D(p)}(pp)
 =2\pi R_{\rm black}\,\delta R(D),
 \nn
\\
&&
\delta R(D)=0.06\pm 0.01\; {\rm fm},
\eea
Recall, $R_{\rm black}\simeq 0.16 \ln(\sqrt{s}/80{\rm GeV})\;{\rm fm}$.

\section{Contribution of $u$-Channel and the
 Real Part of the Scattering Amplitude}

The ratio $Re \,A_{el}/Im\, A_{el}$ at ${\bf q}^2_\perp \simeq 0$ at
asymptotic energies is determined by the analyticity of the
scattering amplitude. Taking into account contributions of $s$ and $u$
channels we write:
\be \label{yf17}
A_{el}\propto i\, [
\ln^2(s/s_0)+\ln^2 (-s/s_0)],\qquad \frac{Re\,
A_{el}}{Im\,A_{el}}\simeq \frac{\pi}{\ln(s/s_{0R})}\,.
\ee
with $-s=se^{-i\pi}$
and
$\ln^2(se^{-i\pi})=\Big(\ln s-i\pi\Big)^2\simeq \ln^2s -2\pi i\ln s$
.

Experimental data ( $Re \,A_{el}/Im\, A_{el}=0.14^{+ 0.01}_{-0.08}$
 \cite{totem} ) require a small value of $s_{0R}$. For 7 TeV Eq.
(\ref{yf17}) gives us $Re \,A_{el}/Im\, A_{el}=0.15$ at
$\sqrt{s_{0R}}=200$ MeV that is an inverse value of the radius
of a black disk at such an energy, $\sim 1$ fm.
So, the $Re \,A_{el}/Im\, A_{el}$
is determined by the size of the disk while the rate of its growth
by the leading pomeron characteristics.

The real part of the scattering amplitude at ${\bf q}^2_\perp > 0$
can be also determined by analyticity requirement. We write
\be \label{yf18}
A_{el}= \frac{i}{2} \bigg[
\ln^2(s/s_0)A(\tau)+\ln^2 (-s/s_0)A(\tau^+)\bigg]
 \, .
\ee
Here
$
\tau=\sigma_{tot}(s){\bf q}^2_\perp$
and
$
\tau^+=\sigma_{tot}(-s){\bf q}^2_\perp \simeq
\Big(1-\frac{2\pi i}{\ln s}\Big)\sigma_{tot}(s)  {\bf q}^2_\perp
= \Big(1-\frac{2\pi i}{\ln s}\Big)\tau
$.
Therefore the scattering amplitude within account of the real part
terms reads:
\be
A_{el}\simeq i\ln^2 s \Big[ A(\tau)-\frac{\pi i}{\ln s} \bigg( A(\tau)+
\tau\frac{dA(\tau)}{d\tau}
\bigg)
\Big] .
\ee

\begin{figure}[h]
\centerline{\epsfig{file=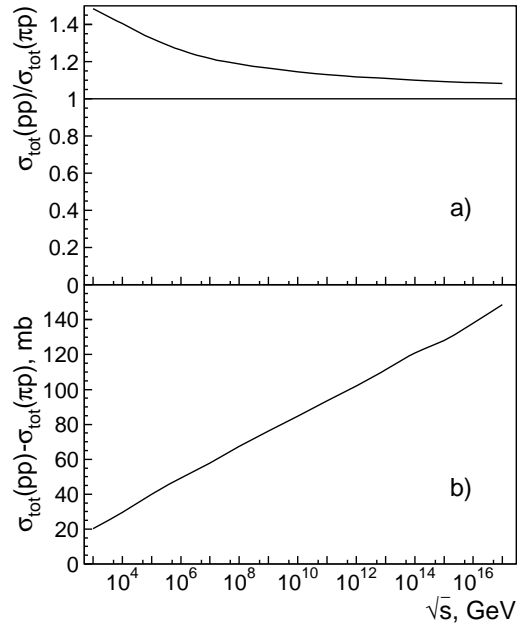,width=7cm}}
\caption{a) Ratio of proton-proton and pion-proton total cross
sections, $\sigma_{tot}(pp)/\sigma_{tot}(\pi p)\to 1$, and
b) its difference $\sigma_{tot}(pp)-\sigma_{tot}(\pi p)\propto \ln s$.
 \label{pi-p-03} }
 \end{figure}


\section{The $\pi p$ and $\pi\pi$ Total Cross Sections
and the Gribov Universality}

The Dakhno-Nikonov model made it possible to extend conventionally
the scheme on $\pi p$ collisions - for that additionally the quark
distribution to the pion is needed only. The distribution of quarks to a
pion is known, see, for example, ref. \refcite{AMN} that allows to give
predictions.

%

\begin{figure*}[ht]
\centerline{
\epsfig{file=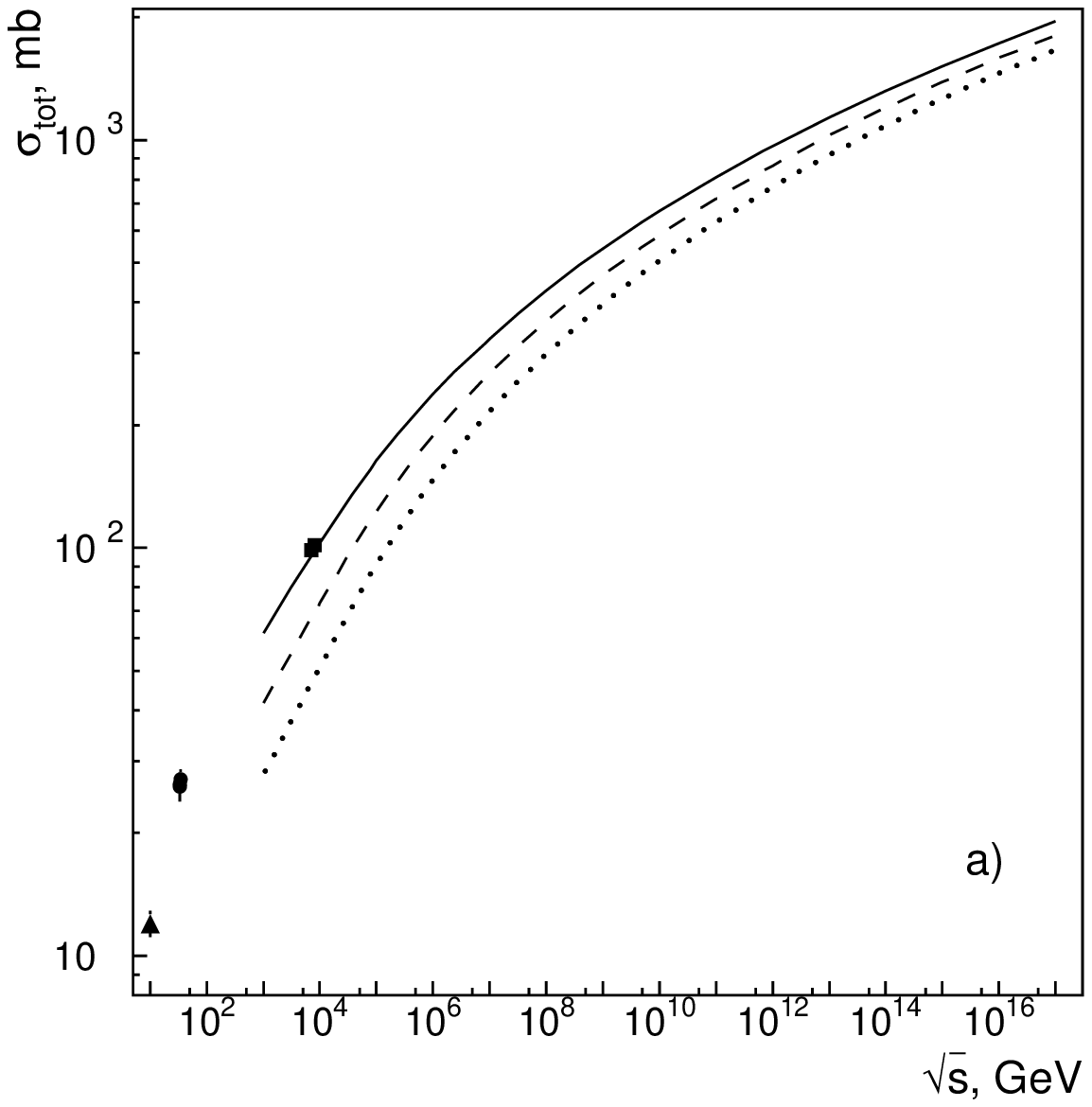,width=0.48\textwidth}
\epsfig{file=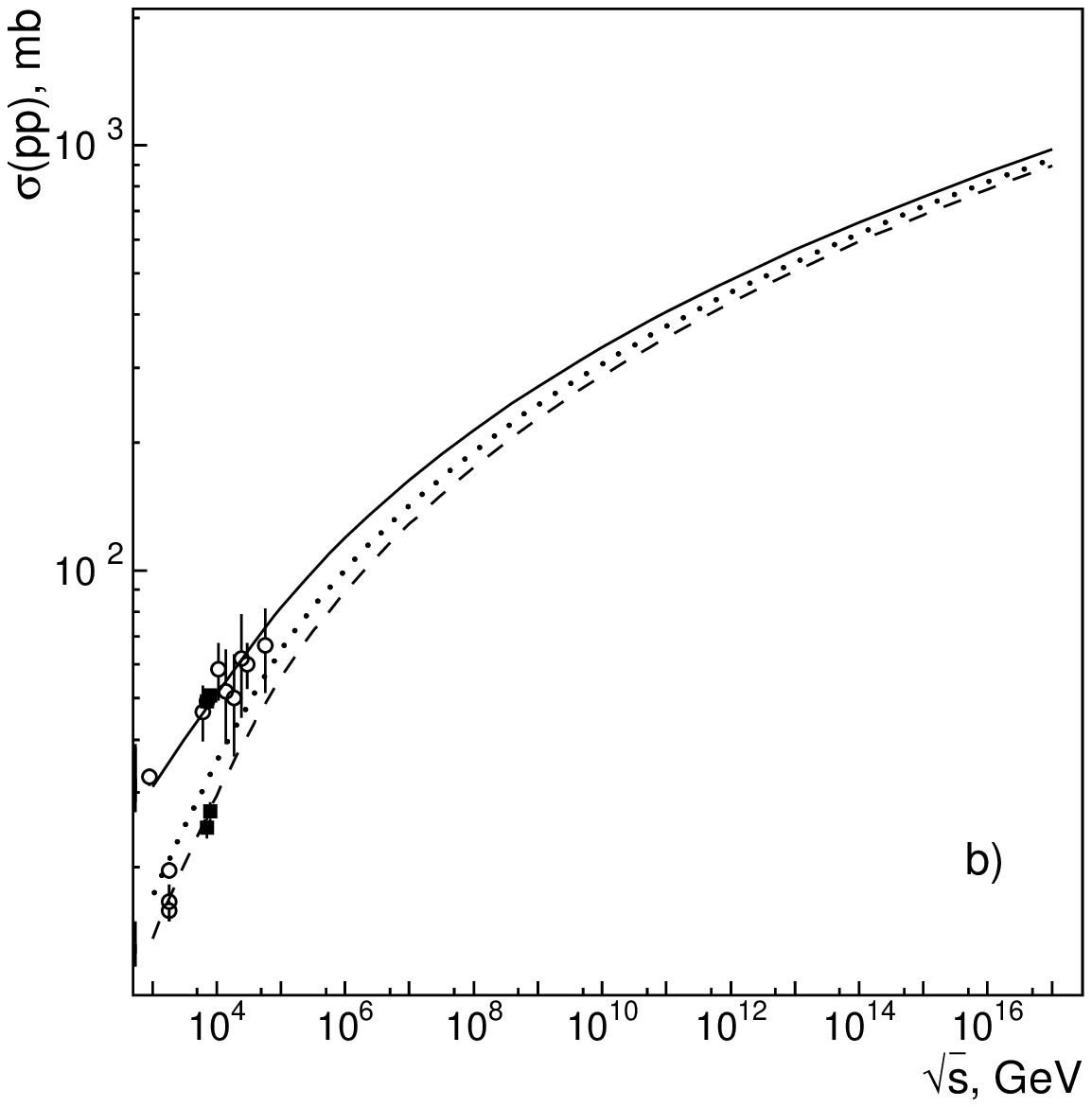,width=0.48\textwidth}}
\caption{ a) Total cross sections: $\sigma_{tot}(pp)$ (solid line),
$\sigma_{tot}(\pi p)$ (dashed line) and
$\sigma_{tot}(\pi \pi)$ (dotted line).
 Squares ($pp$) are from
\cite{totem},
circles ($\pi p$) are from \cite{Dersch:1999zg},
triangles ($\pi \pi$) are from \cite{Abramowicz:1979ca}.
b) Proton-proton
collisions: $\frac12\sigma_{tot}(pp)$ (solid line), $\sigma_{el}(pp)
+2\sigma_{pD}(pp)$ (dotted line) and $\sigma_{el}(pp)$ (dashed line).
Data \cite{pre1,pre2,pre3,pre4,pre5,totem,auger} stand for
$\frac12\sigma_{tot}(pp)$ and $\sigma_{el}(pp)$.} \label{f11}
\end{figure*}

The ratio $\sigma_{tot}(pp)/\sigma_{tot}(\pi p)$ is shown in
Fig.~\ref{pi-p-03}a, it asymptotically tends to 1. The
difference of proton-proton and pion-proton total cross sections is
increasing with energy, Fig.~\ref{pi-p-03}b, and can be
described at $\sqrt{s}\ga 10^6$ GeV as
 \be
\sigma_{tot}(pp)-\sigma_{tot}(\pi p)\simeq
1.9\ \ln\frac{s}{s_0}\; {\rm mb}, \qquad s_0=0.8 \, {\rm GeV}^2.
 \label{yf19}
 \ee
 The universality of the total cross sections means the equality of
the leading terms ($\propto \ln^2 s$) only, and it is demonstrated
in Fig.~\ref{f11}a where $\sigma_{tot}(pp)$, $\sigma_{tot}(\pi p)$
and $\sigma_{tot}(\pi \pi)$ are shown.

%

\section{Conclusion}

The twofold structure of hadrons -- hadrons are built by constituent
quarks and the latter are formed by clouds of partons -- manifests
itself in hadron collisions. At moderately high energies colliding
protons reveal themselves in the impact parameter space as three
disks corresponding to three constituent quarks, Fig.~\ref{disks}a.
At ultra-high energies the situation is transformed to a one-disk
picture, Fig.~\ref{disks}c, and the energy of this transformation is
that of LHC. The radius of the black disk at asymptotic energies is
increasing,  $R_{\rm black}\simeq 2 \sqrt{\Delta\alpha'_P}\ln s$,
that provides a $\ln^2s$ growth of total and elastic cross sections
$\sigma_{tot}\simeq 2\pi R^2_{\rm black}$ and
$\sigma_{el}\simeq \pi R^2_{\rm black}$. Diffractive dissociation
cross sections are increasing as $\ln s$ being related to the black disk
rim effect: $\sigma_D \propto 2\pi R_{\rm black}$ and
$\sigma_{DD}\propto 2\pi R _{\rm black}$.

A steady increase of the black disk radius $R_{\rm black}$,
is determined by parameters of the leading $t$-channel singularity only:
that are the pomeron intercept $\alpha(0)=1+\Delta$ ($\Delta>0$) and the
pomeron trajectory slope $\alpha'_P$. The $s$-channel unitarization
of the scattering amplitude damps the strong pomeron pole
singularity transforming it into a multipomeron one. Therefore, we
face the intersection of problems of the gluon content of the
$t$-channel states at ultrahigh energies and the physics of gluonic
states, glueballs. At present the glueball states are subjects of
intensive investigations, see, for example
\refcite{vva-usp,book3,ochs} and references therein.  Studies
of phenomena related to glueballs and multigluon states at small
$|t|$ are enlightening for the confinement
singularity problem - see discussion in ref. \refcite{conf-PR}. The
large value of mass of the soft effective gluon (and the corresponding
values of the low-lying glueballs) and the slow rate of the black disk
increase appear to be closely related phenomena.

The change of the regime, from the constituent quark collision
picture to that with a united single disk, was discussed in
\refcite{ani-she,ani-lev-rys,ani-bra-sha} when definite indications
about the hadron cross section growth appears. It was emphasized that the
single black disk regime should change probabilities of productions
of hadrons in the fragmentation region (hadrons with
$x=p/p_{in}\sim 1$), for more detailed discussion see~\refcite{book2}.


\section*{Acknowledgment}
We thank Ya.I. Azimov, M.G. Ryskin,
and A.V. Sarantsev for useful discussions and comments.
 The work was supported by grants RFBR-13-02-00425 and
 RSGSS-4801.2012.2.



\begin{thebibliography}{99} \fussy

\bibitem{pre1}
UA4 Collaboration,   Phys. Lett. {\bf B147} (1984)  385.
\bibitem{pre2}
UA4/2 Collaboration, Phys. Lett. {\bf B316} (1993)  448.
\bibitem{pre3}
UA1 Collaboration,   Phys. Lett. {\bf B128} (1982)  336.
\bibitem{pre4}
E710 Collaboration,  Phys. Lett. {\bf B247} (1990)  127.
\bibitem{pre5}
CDF Collaboration,   Phys. Rev.  {\bf D50}  (1994) 5518.




\bibitem{Froi} M. Froissart, Phys. Rev. {\bf 123} (1961) 1053.

\bibitem{Kaid}A.B. Kaidalov and K.A. Ter-Martirosyan,
Sov. J. Nucl. Phys. {\bf 39} (1984) 979.
\bibitem{Land}  A. Donnachie and P.V. Landshoff, Nucl. Phys. {\bf B231}
(1984) 189.
\bibitem{_4glauber1}R.J. Glauber, Phys. Rev. {\bf 100} (1955) 242.
\bibitem{_4glauber2}R.J. Glauber,
{\em Lectures in Theoretical Physics}, ed. W.E. Britten, L.G. Danham, Vol.1,
New York (1959) 315.


\bibitem{Gaisser}  T.K. Gaisser and T. Stanev, Phys. Lett.,
{\bf B219} (1989) 375.

\bibitem{Block} M. Block, F. Halzen and B. Margolis, Phys.
Lett., {\bf B252}  (1990) 481.

\bibitem{Fletcher}R.S. Fletcher, Phys. Rev. {\bf D46} (1992) 187.

\bibitem{azimov} Y.I. Azimov, Phys. Rev. {\bf D84}, 056012 (2011);
arXiv:1208.4304.

\bibitem{atlas} ATLAS collaboration, Nature Commun. {\bf 2}
(2011) 463.

\bibitem{cms}  CMS collaboration, Phys. Lett. {\bf B722} (2013) 5.

\bibitem{totem} G. Latino for the TOTEM collaboration,
{\it Summary of Physics Results from the\\ TOTEM Experiment},
 arXiv:1302.2098.

\bibitem{auger} Pierre Auger Collaboration (P. Abreu et al.),
Phys. Rev. Lett. {\bf 109} (2012) 062002.

\bibitem{sch-rys}V.A. Schegelsky, M.G. Ryskin, Phys. Rev. {\bf D85},
 (2012) 094024.


\bibitem{block}M.M. Block and F. Halzen,
Phys. Rev.  {\bf D86} (2012) 051504.


\bibitem{ryskin} M.G. Ryskin, A.D. Martin and V.A. Khoze,
Eur. Phys. J.  {\bf C72} (2012) 1937.

\bibitem{dremin1} I.M. Dremin and A.A. Radovskaya,
Europhys. Lett. {\bf 100} (2012) 61001.



\bibitem{martynov} E. Martynov,
  Phys. Rev. {\bf D87} (2013) 114018.

\bibitem{Ryskin:2009tj}
M.~Ryskin, A.~Martin, and V.~Khoze,
Eur.Phys.J. {\bf C60} (2009) 249.

\bibitem{KMR-s3}
M.~Ryskin, A.~Martin, and V.~Khoze,
Eur.Phys.J. {\bf C71} (2011) 1617.

\bibitem{KMRLHC}
V.~Khoze, A.~Martin, and M.~Ryskin,
Eur.Phys.J. {\bf C73} (2013) 2503.

\bibitem{Khoze:2013jsa}
V.~Khoze, A.~Martin, and M.~Ryskin,
arXiv:1312.3851.

\bibitem{Maor}
U.~Maor,
arXiv:1310.7340.

\bibitem{Gotsman:2013lya}
E.~Gotsman,
arXiv:1304.7627.

\bibitem{Ost}
S.~Ostapchenko,
Phys.Rev. {\bf D81} (2010) 114028.


\bibitem{DN} L.G. Dakhno and V.A. Nikonov,
Eur. Phys. J. {\bf A8} (1999) 209.

\bibitem{GW}M.L. Good, W.D. Walker, Phys. Rev. {\bf 120}
(1960) 1857.



\bibitem{ann1}
V.V. Anisovich, K.V. Nikonov and V.A. Nikonov,
Phys. Rev.  {\bf D88} (2013) 014039.

\bibitem{ann2}
V.V. Anisovich, V.A. Nikonov and J. Nyiri
Phys. Rev.  {\bf D88} (2013) 094015.



\bibitem{block3}M.M. Block and F. Halzen,
  arXiv:1210.4086v1.


\bibitem{1212.5096}
M.J. Menon and P.V.R.G. Silva,  arXiv:1212.5096v1.

\bibitem{gribov-tot}V.N. Gribov, Yad. Fiz. {\bf 17} (1973) 603,
[Sov. J. Nucl. Phys. {\bf 17} (1973) 313].



\bibitem{ani-she}V.V. Anisovich and V.M. Shekhter, Yad. Fiz.
{\bf 28} (1978) 1079,
[Sov. J. Nucl. Phys. {\bf 28} (1978) 554].

\bibitem{ani-lev-rys}V.V. Anisovich, E.M. Levin and M.G. Ryskin,
Yad. Fiz. {\bf 29}, 1311 (1979),
[Sov. J. Nucl. Phys. {\bf 29}, 674 (1979)].

\bibitem{ani-bra-sha}V.V. Anisovich, V.M. Braun and Yu.M. Shabelski,
Yad. Fiz. {\bf 36} (1982) 1556,
[Sov. J. Nucl. Phys. {\bf 36} (1982) 904].

\bibitem{book2}V.V. Anisovich, M.N. Kobrinsky, J. Nyiri, Yu.M.
Shabelski, {\it Quark model and high energy collisions}, Second
  Edition, World Scientific, Singapore (2004).

\bibitem{vva-usp}V.V. Anisovich, Usp. Fiz. Nauk {\bf 47} (2004) 45
[Phys. Usp. {\bf 47} (2004) 45].

\bibitem{book3}
A.V. Anisovich, V.V. Anisovich, M.A. Matveev, V.A. Nikonov, J. Nyiri,
A.V. Sarantsev,
{\it Mesons and Baryons. Systematization and Methods of Analysis},
World Scientific, Singapore (2004).


\bibitem{klempt-zaitsev} E. Klempt and A. Zaitsev, Phys. Rept.
{\bf 454} (2007) 1.
\bibitem{ochs} W. Ochs, J. Phys. {\bf 40} (2013) 043001.
\bibitem{conf-PR}
A.V. Anisovich, V.A. Nikonov,
A.V. Sarantsev, V.V. Anisovich, M.A. Matveev, T.O. Vulfs, J. Nyiri,
Phys. Rev. {\bf D84} (2011) 076001.

\bibitem{adnPR} V.V. Anisovich, L.G. Dakhno, and V.A. Nikonov, Phys.
Rev. D{\bf 44}, 1385 (1991).



%




\bibitem{AMN}
 V.V. Anisovich,    D.I. Melikhov, and V.A. Nikonov,
     Phys. Rev.  {\bf D52} (1995) 5295.

\bibitem{DL}A. Donnachie and P.V. Landshoff,
arXiv:11122485, arXiv:1309.1292.


















\bibitem{uzhi}V. Uzhinsky and A. Galoyan,
 arXiv:1111.4984v5.


\bibitem{dremin2} I.M. Dremin, Nucl. Phys. {\bf A888} (2012) 1.



%
%
%
%
%
%
%
%
%
%
%
%
%
%
%
%
%
%
%
%
%


\bibitem{Dersch:1999zg}
  U.~Dersch {\it et al.}  [SELEX Collaboration],
  Nucl.\ Phys.\  {\bf B579} (2000) 277.

\bibitem{Abramowicz:1979ca}
  H.~Abramowicz, M.~Gorski, G.~Sinapius, A.~Wroblewski, A.~Zieminski,
   H.~J.~Lubatti, K.~Moriyasu and C.~D.~Rees {\it et al.},
  Nucl.\ Phys.\ {\bf B 166} (1980) 62.



\end{thebibliography}
   \end{document}